\documentstyle[twocolumn,aas2pp4]{article}
\def\stacksymbols #1#2#3#4{\def\theguybelow{#2}
	\def\verticalposition{\lower#3pt}
	\def\spacingwithinsymbol{\baselineskip0pt\lineskip#4pt}
	\mathrel{\mathpalette\intermediary#1}}
\def\intermediary #1#2{\verticalposition\vbox{\spacingwithinsymbol
	\everycr={}\tabskip0pt
	\halign{$\mathsurround0pt#1\hfil##\hfil$\crcr#2\crcr
		\theguybelow\crcr}}}

\def\gta{\stacksymbols{>}{\sim}{3}{.5}}

\begin{document}
%

\title{COSMOLOGICAL AND ENVIRONMENTAL INFLUENCES ON 
HOT GAS OBSERVED IN ELLIPTICAL GALAXIES$^1$}


\author{William G. Mathews$^2$ and Fabrizio Brighenti$^{2,3}$}

\affil{$^2$University of California Observatories/Lick Observatory,
Board of Studies in Astronomy and Astrophysics,
University of California, Santa Cruz, CA 95064\\
mathews@lick.ucsc.edu}

\affil{$^3$Dipartimento di Astronomia,
Universit\`a di Bologna,
via Zamboni 33,
Bologna 40126, Italy\\
brighenti@astbo3.bo.astro.it}






\vskip .2in

\begin{abstract}

The X-ray emission from hot gas in 
bright elliptical galaxies often extends 
far beyond the radius of the stellar system.
This ``circumgalactic'' gas accounts 
for most or all of 
the large spread in X-ray luminosity $L_x$ 
among ellipticals having similar optical luminosities $L_B$.

We have developed gas dynamical models 
describing the evolution of 
gas within and around ellipticals beginning 
with an overdensity perturbation in 
a simple flat cosmology.
At some early time 
we form the stellar galaxy and release supernova energy,
conserving dark and baryonic matter.
We follow the subsequent evolution of intergalactic 
and interstellar gas to the present time.
These models confirm 
that hot gas density and temperature 
distributions currently observed in massive, group-dominant 
ellipticals can be 
understood as a combination of intergalactic gas 
that has flowed into the galaxy group over time 
with gas lost from galactic stars.
Furthermore, if the hot gas and dark matter halos are 
subject to differential 
tidal truncations or mass exchanges between 
group members, then the observed correlation between 
$L_x/L_B$ 
and the relative sizes of galactic 
X-ray images can be understood.
The distribution and physical properties of 
hot interstellar gas observed in massive ellipticals today 
are sensitive to the cosmic baryon fraction, 
the time of maximum star formation and the 
amount of ``feedback'' energy delivered to the 
gas by Type II supernovae at the epoch of galaxy formation.

\end{abstract}





For many years X-ray observers have noticed that 
the X-ray luminosities of elliptical galaxies span a huge 
range for galaxies of similar optical luminosity $L_B$ 
(Fabbiano 1989; Eskridge, Fabbiano, \& Kim 1995). 
Attempts to find correlations between $L_x/L_B$ and 
other galactic parameters had been largely unsuccessful.
However, in a recent compilation of {\it ROSAT} data, 
Mathews \& Brighenti (1998) discovered that 
$L_x/L_B$ correlates strongly with the size of the X-ray 
image $r_{ex}/r_e$, where $r_{ex}$ is the projected radius 
that contains half the total $L_x$ and $r_e$ is the optical 
half-light radius.
A correlation between $L_x/L_B$ and $r_{ex}/r_e$ was suspected 
because the hot gas in well-resolved, X-ray luminous 
ellipticals often extends far beyond the optical
images where it could be rather stochastically influenced by tidal 
mass exchanges. 

The very extended X-ray halos around many bright ellipticals 
also indicate that most of this gas is unlikely to have 
been ejected by old galactic stars during their post-main sequence 
evolution.
Furthermore, the average gas temperature in 
bright ellipticals exceeds the equivalent stellar 
temperature $T_*$ by $\sim 1.5$ (Davis \& White 1996), 
suggesting again that the hot gas is in virial equilibrium
in the more massive 
and more extended dark halos surrounding the galaxies. 
Using detailed hydrodynamic models, 
Brighenti \& Mathews (1998) showed that 
the extended hot gas density and temperature 
profiles in X-ray bright ellipticals can 
be understood only if 
most of the hot interstellar gas is very old, dating 
from the epoch of galaxy formation, and has not come solely 
from mass loss from galactic stars.

In this letter we combine these two lines of inquiry. 
First, using a simple cosmological model we show that 
the additional ``circumgalactic'' gas required to
account for large values of $L_x/L_B$ can result quite naturally 
from secondary infalling intergalactic gas that has accumulated in 
the outer dark halo over a Hubble time.
Second, we demonstrate that the correlation between 
$L_x/L_B$ and $r_{ex}/r_e$ can indeed be generated by tidal 
truncations of these same models.

For a fully self-consistent model of 
the global evolution of gas in 
ellipticals, it is necessary to begin with 
a galaxy-group sized perturbation in an expanding 
universe, allow for the formation of the stellar elliptical, 
and include the release of supernova energy 
that accompanies early star formation.
For simplicity we consider here a localized 
density perturbation in 
a simple flat cosmology with various baryonic 
fractions $\Omega_b/\Omega$.
After a few gigayears, when enough baryonic matter has 
accumulated to create the giant elliptical, we construct 
the stellar galaxy with a de Vaucouleurs profile
and simultaneously release the energy from Type II supernovae.
By forming the galaxy in this instantaneous fashion,
we circumvent the dynamically complex merging processes 
that must have occurred (e. g. Merritt 1985).
After the stellar system is formed, baryonic and dark 
matter continue to flow into the galactic perturbation.
The baryonic matter shocks and compresses to approximately the virial 
temperature in the halo, $T \sim 10^7$ K, similar to interstellar 
gas temperatures observed today.
With this simple model we conserve baryonic and dark matter
and allow properly for the influence of supernova energy.
We study the subsequent gas dynamical evolution of the 
intergalactic and 
interstellar gas in detail, allowing for stellar mass loss 
and Type Ia supernovae. 
The objective in this calculation 
is to reproduce the density and temperature 
profiles observed in the hot gas within and around bright 
ellipticals today.
A more detailed discussion of our hydrodynamical models  
has been submitted to the Astrophysical Journal.

\vskip.2in

At the present time only about a dozen bright, nearby ellipticals
have been spatially resolved in soft X-rays.
For these galaxies, 
the X-ray surface brightness and projected spectral variation 
can be inverted to determine the electron density $n(r)$ 
and temperature $T(r)$ profiles with physical radius.
Assuming approximate hydrostatic equilibrium,
the density and temperature gradients can be used to 
determine the total internal mass $M(r)$
(Brighenti \& Mathews 1997).
As a guide in evaluating the success of the models we discuss here, 
we choose NGC 4472, a typical bright 
E2 elliptical in Virgo that is the dominant elliptical in a 
local galaxy subgroup. 
Data for NGC 4472 is available from {\it Einstein} HRI 
(Trinchieri, Fabbiano, \& Canizares 1986) and 
{\it ROSAT} HRI and PSPC (Irwin \& Sarazin 1996).
The interstellar gas density profile for NGC 4472 has been 
determined out to almost $17 r_e \approx 150$ kpc 
(at a distance of 17 Mpc).
NGC 4472 appears to be colliding with a small 
dwarf galaxy and interacting with more extended Virgo 
intercluster gas (Irwin \& Sarazin 1996; Irwin, Frayer, \& Sarazin 
1997).
This may explain the azimuthal
asymmetry in the X-ray image beyond $\sim 3.5$ kpc. 
In spite of this undesirable asymmetry, in these times of 
pre-AXAF resolution NGC 4472 is one of the 
few normal bright ellipticals in which the stellar component 
is clearly seen in the mass profile determined from X-ray
observations; 
this is useful for our models to accurately evaluate the 
stellar potential, the gas contributed by stellar mass loss, etc.
The X-ray profile in NGC 4649 is also sensitive to the stellar 
potential and its X-ray properties are very similar to those 
of NGC 4472 (Brighenti \& Mathews 1997), 
but X-ray observations in NGC 4649 are not as spatially 
extensive as in NGC 4472.
On balance, we adopt 
the global, azimuthally averaged X-ray properties of 
NGC 4472 as generally representative of other 
X-ray bright ellipticals.
 
The total mass distribution for NGC 4472 found by Brighenti 
\& Mathews (1997) can be fit quite well with a 
de Vaucouleurs stellar core out to $\sim r_e$ and an 
Navarro-Frenk-White halo (NFW) 
beyond (Navarro, Frenk, \& White 1996).
In making these fits we use a total stellar mass of 
$M_* = 7.26 \times 10^{11}$ $M_{\odot}$ which is based 
on $L_B = 7.89 \times 10^{10}$
$L_{B,\odot}$ and the mass to light ratio 
$M_*/L_B = 9.20$ of van der Marel (1991); this 
$M_*/L_B$ is in fact verified by the X-ray image of NGC 4472
(Brighenti \& Mathews 1997). 
For the one-parameter NFW halo we use a virial mass of 
$M_h = 4 \times 10^{13}$ $M_{\odot}$ which corresponds 
to an NFW concentration parameter of $c = 10.47$. 
The total mass distribution, stars plus dark halo, agrees 
very well with the X-ray determined $M(r)$, but is slightly 
too massive near $\sim r_e$ suggesting that the dark halo
may be less centrally peaked than NFW predict.

In our hydrodynamical model for the evolution of the 
intergalactic and interstellar gas, we adopt 
a very simple flat cosmology with $\Omega = 1$ and 
$H = 50$ km s$^{-1}$ Mpc$^{-1}$, corresponding to a current universal 
age of $t_n = 13$ Gyrs.
Primordial nucleosynthesis restricts the baryonic 
mass fraction to a narrow range,
$\Omega_b (H/50)^2 \approx 0.05 \pm 0.01$ (Walker et al. 1991).
We assume that 
both baryonic and dark matter converge toward the 
proto-4472 galaxy perturbation as described 
by the self-similar solution of Bertschinger (1985).
In this model all flow parameters depend on 
a single similarity variable $\lambda = r/r_{ta}(t)$
where $r_{ta}(t) \propto t^{8/9}$ 
is the turnaround radius where 
the cosmic flow velocity vanishes at time $t$.

Before we form the galaxy at time $t_*$ the cold 
baryonic gas within $r_{ta}(t)$ 
flows exactly with the dark matter until 
it encounters an accretion shock. 
The time-dependent 
dark matter potential is described 
by an outer Bertschinger flow, similar to that into a 
central singularity, and an inner stationary NFW halo 
characterized by parameters that fit $M(r)$ for NGC 4472.
At each time the NFW and Bertschinger mass distributions
are matched at some radius $r_c(t)$ that conserves 
the total mass of dark matter.
Outside both $r_c(t)$ and the accretion shock, 
the mass flux in the baryonic component at fixed radius
varies as $\rho u \propto t^{-1/3}$ so most of the halo gas 
is accumulated at early times.
The accretion
shock forms at a very large radius, 
$r_{sh} \gta 1$ Mpc for $t \gta 6$ Gyrs.
Within the accretion shock a cooling flow is is established 
and the gas velocity is very subsonic.

At some time $t_*$ when a sufficient number of baryons has 
collected in the center of the flow, we form the 
de Vaucouleurs stellar 
mass distribution $M_*(r)$ matching NGC 4472.
This is done by reducing the gas density within the 
accretion shock radius $r_{sh}(t_*)$, lowering its mass by $M_*$.
Also at this time the 
remaining gas in $r < r_{sh}(t_*)$ is heated
by the Type II supernova energy from massive stars,
$m > 8$ $M_{\odot}$.
We assume a Salpeter power law IMF from $m_{\ell} = 0.08$
to $m_u = 100$ $M_{\odot}$ for which 
$\eta_{II} = 6.81 \times 10^{-3}$ SNII are produced 
per $M_{\odot}$ of stars formed, each of energy 
$E_{sn} = 10^{51}$ ergs.
The amount of SNII energy received by the hot gas depends 
both on the uncertain IMF-dependent value $\eta_{II}$ 
and the efficiency that this energy is transferred to the 
hot gas and not lost by radiation.
For this reason we 
adopt $\eta_{std} E_{sn} M_*$ with 
$\eta_{std} = 6.81 \times 10^{-3}$ 
as a reference energy which can be adjusted as needed.

After time $t_*$ the stars lose mass as described by 
$\alpha_*(t) = 4.7 \times 10^{-20} [t/(t_s)]^{-s}$ s$^{-1}$
where $t_n = 13$ Gyrs is the current time, $t_s = t_n - t_*$ 
and $s = 1.26$.
We assume that the gas is heated after $t_*$ by Type Ia 
supernova, each of energy $E_{sn}$, at a rate 
${\rm SNu}(t) = {\rm SNu}(t_n) (t / t_n)^{-p}$ 
with $p = 1$ and SNu$(t_n) = 0.03$ supernovae per 100 years
per $10^{10} L_{B,\odot}$.

The gas dynamical equations we solve are 
$${ \partial \rho \over \partial t}
+ {1 \over r^2} { \partial \over \partial r}
\left( r^2 \rho u \right) = \alpha \rho_*,$$
$$\rho \left( { \partial u \over \partial t}
+ u { \partial u \over \partial r} \right)
= - { \partial P \over \partial r}
- \rho {G M_{tot}(r,t) \over r^2} - \alpha \rho_* u,$$
and
$$ \rho {d \varepsilon \over dt } =
{P \over \rho} {d \rho \over d t}
- { \rho^2 \Lambda \over m_p^2}
+ \alpha \rho_*
\left[ \varepsilon_o - \varepsilon - {P \over \rho}
+ {u^2 \over 2} \right]$$
where $\varepsilon = 3 k T / 2 \mu m_p$ is the
specific thermal energy;
$\mu = 0.62$ is the molecular weight and $m_p$ the proton mass.
$M_{tot}(r,t)$ is the total mass of stars, dark matter
and hot gas within radius $r$.
The mean injection energy of gas from stars is
$\varepsilon_o = 3 k T_o / 2 \mu m_p$ where
$T_o = (\alpha_* T_* + \alpha_{sn} T_{sn})/\alpha$
and $\alpha = \alpha_* + \alpha_{sn}$.
$\Lambda(T)$ is the coefficient for 
optically thin radiative cooling.
These equations are described more fully by 
Brighenti \& Mathews (1998).
We solve them numerically 
using a spherical 1D Eulerian hydrocode (ZEUS) with
logarithmic spatial zoning.

\vskip.2in

In Figure 1 we illustrate the results of several gas-dynamical
calculations at time $t_n = 13$ Gyrs using 
$t_* = 2$ Gyrs, $\eta_{II} = \eta_{std}$ and 
$\Omega_b = 0.05$ or 0.06.
Considering the relative simplicity of our model, 
the agreement with both the temperature and density 
distributions observed in NGC 4472 is excellent.
The positive gas temperature gradient within $r \sim 3 - 5 r_e$ 
kpc observed in NGC 4472 and other bright ellipticals
(Brighenti \& Mathews 1997) occurs as hot 
inflowing halo gas mixes 
with the slightly cooler gas at $\sim T_*$ 
that characterizes gas ejected from stars orbiting 
within a few $r_e$.
The apparent sensitivity of our results to $\Omega_b$ is
of considerable interest.
However, we emphasize that the influence of the parameters
$\Omega_b$, $\eta_{II}$ and $t_*$ is degenerate.
For example, solutions with $\Omega_b$ decreased by $\sim 0.01$
are similar to those with $\eta_{II}$ increased by $\sim 2$ or 
with $t_*$ decreased by $\sim 1$ Gyr.
This degeneracy can be removed with more detailed models 
that include the abundance of iron and other elements produced 
in supernovae (Renzini 1998).
Nevertheless, the $\Omega_b = 0$ 
solution shown in Figure 1, 
for which the only source of gas since $t_*$ 
is stellar mass loss, is clearly inadequate.
We conclude that cosmic inflow of gas is essential 
in understanding 
the properties of hot gas observed in giant ellipticals today.
Moreover, this extended ``circumgalactic'' gas is sensitive 
to important cosmological parameters and to the details of 
SNII energy release during the epoch of galaxy formation;
solutions with little or no SNII energy provide very poor 
fits in Figure 1.

In Figure 2 we show the current 
distribution of the baryon fraction 
$f_b = \Omega_b/\Omega$ with radius for the 
$\Omega_b = 0.05$, $\eta_{II} = \eta_{std}$ solution.
The baryonic cavitation around the 
galaxy is a relic of the SNII energy released at $t_*$ 
when gas was pushed out of the central parts of the flow.
This suggests that determinations of $f_b$ 
based on X-ray observations in the range 
$\sim 100 - 300$ kpc could either underestimate 
or overestimate the true cosmic value of $\Omega_b$. 
Obviously, the variation of $f_b(r)$ in $r \gta 300$ kpc 
shown in Figure 2 cannot 
be relevant to NGC 4472 because of the gravitational 
influence of the Virgo cluster.

As a further confirmation of the overall success of 
our models, we compare in Figure 3 the locus of truncated models 
with observed ellipticals
in the ($L_x/L_B,~r_{ex}/r_e$)-plane.
The large circle, cross and square at the right are the current 
loci of untruncated models with $\Omega_b 
= 0.04$, 0.05 and 0.06; we have made these models somewhat more
gas rich than NGC 4472 by reducing $\eta_{II}$. 
When these models are truncated at time $t_{tr} = 9$ Gyrs 
the current position of the galaxy at time $t_n = 13$ Gyrs 
moves progressively along the observed sequence of ellipticals 
as the truncation radius $r_{tr}$ decreases from 500 kpc 
to 100 kpc.
At the time of truncation, all gas and dark matter is removed 
at $r > r_{tr}$, causing a rarefaction wave to propagate 
into the galactic ISM; 
the cosmic inflow is stopped after $t_{tr}$
and a new equilibrium is established 
in a sound crossing time $\sim 1$ Gyr.
These results are insensitive to the time $t_{tr}$ when the
truncation occurred as long as $t_{tr} \gta 6$ Gyrs.
This insensitivity to $t_{tr}$ arises from the 
rather small changes in the gas flow variables 
within the accretion shock until the present time, 
the large radius of the accretion shock at early times
($r_{sh} \approx 400$ kpc at $t = 4$ Gyrs),
and the declining rate of intergalactic gas accumulation 
with time.
The density and temperature profiles at $t_n$ for the 
truncation solution that best matches NGC 4472
($\Omega_b = 0.05$, $r_{tr} = 400$ kpc) are shown in Figure 1.
This suggests that NGC 4472 may have been tidally truncated
in the past.
Finally, we expect that purely gaseous truncations produced by 
ram pressure stripping in dense clusters 
would also reduce both $L_x/L_B$ and $r_{ex}/r_e$ 
(Sakelliou \& Merrifield 1998).

\vskip.2in

\noindent
Our main conclusions:

\noindent
${\bullet}$ The extended ``circumgalactic'' hot gas surrounding massive 
ellipticals can be understood as a natural consequence of 
inflowing intergalactic gas over the Hubble time.
This gas cannot have been produced by mass loss from stars 
within the elliptical itself.

\noindent
${\bullet}$ Gas provided by stellar mass loss virializes to a relatively 
lower temperature $\sim T_*$ and, when mixed with hotter gas flowing 
in from the outer halo, creates the positive gas 
temperature gradients commonly observed within $\sim 3r_e$
(Brighenti \& Mathews 1997).

\noindent
${\bullet}$ Currently observed hot gas density and temperature profiles 
are sensitive to $\Omega_b$, to the time of galaxy formation, 
and to the total ``feedback'' energy 
released by Type II supernovae as galactic stars formed. 

\noindent
${\bullet}$ Because of the SNII-driven outflow just after 
the time of galaxy formation,
the cosmic baryonic fraction $\Omega_b/\Omega$ 
cannot be accurately measured 
from X-ray observations or dynamical arguments within several
hundred kpc.

\noindent
${\bullet}$ The fraction of hot gas currently within $\sim r_e$ 
that comes from stellar mass loss is  
strongly dependent on $\Omega_b$ and the SNII energy.
This must be considered in interpreting the abundance
of iron and other elements in the hot gas.

\noindent
${\bullet}$ The enormous range in $L_x/L_B$ among elliptical galaxies
is due in large part to differential truncation of the outer halos.
The truncation could be caused by massive nearby
group member galaxies or by another nearby group or cluster.
It is likely that NGC 4472 has been truncated,
perhaps by nearby group companions or by the Virgo cluster.
If the wide range in $L_x/L_B$ were due to 
Type Ia supernova-driven galactic winds,
as is often suggested, the $L_x/L_B,~r_{ex}/r_e$ correlation
would be difficult to explain and the 
hot gas iron abundance would depend on $L_x/L_B$.

\acknowledgments

Our work on the evolution of hot gas in ellipticals is supported by
NASA grant NAG 5-3060 for which we are very grateful. In addition
FB is supported 
in part by Grant ARS-96-70 from the Agenzia Spaziale Italiana.






\vskip.1in
\figcaption[aasrexfig1.ps]{
Gas density ({\it Upper panel}) and temperature 
({\it Lower panel}) of several models at $t_n = 13$ Gyrs 
compared to observations of NGC 4472.
{\it Solid line:} $\Omega_b = 0.05$, $t_* = 2$ Gyrs, $\eta_{II} =
\eta_{std}$;
{\it Dashed line:} $\Omega_b = 0.06$, $t_* = 2$ Gyrs, $\eta_{II} =
\eta_{std}$;
{\it Short dashed line:} $\Omega_b = 0.05$, $t_* = 2$ Gyrs, $\eta_{II} =
0.3 \eta_{std}$, $t_{tr} = 9$ Gyrs, $r_{tr} = 400$ kpc;
{\it Dash-dotted line:} $\Omega_b = 0.00$, $t_* = 2$ Gyrs, 
$\eta_{II} = 0$.
Filled circles are {\it Einstein HRI} gas density 
observations from Trinchieri, Fabbiano \& Canizares (1986); 
open circles 
are {\it ROSAT} density and temperature 
observations from Irwin \& Sarazin (1996).
\label{fig1}}

\vskip.1in
\figcaption[aasrexfig2.ps]{
{\it Solid line:} Variation of the baryon fraction 
$f_b$ at time $t_n = 13$ Gyrs.
{\it Dashed line:} The baryonic contribution of the gas alone. 
The model parameters are $\Omega_b = 0.05$, $t_* = 2$ Gyrs, 
and $\eta_{II} = \eta_{std}$.
At very large radii the baryonic fraction approaches 
$\Omega_b = 0.05$.
\label{fig2}}

\vskip.1in
\figcaption[aasrexfig3.ps]{
A plot of the
($L_x/L_B,~r_{ex}/r_e$)-plane showing the correlation 
among observed ellipticals taken from Mathews \& Brighenti (1998).
The large circle, cross and square at the right are 
loci of untruncated, gas-rich models at $t_n$ 
with the following parameters:
($\Omega_b$, $t_*$, $\eta_{II}$) =  
(0.04, 3, 0.1), (0.05, 2, 0.3), and (0.06, 2, 1) respectively.
The smaller symbols that extend from the untruncated 
solution toward the lower left show the current ($t = t_n$) loci  
of models truncated at time $t_{tr} = 9$ Gyrs 
at five decreasing truncation radii $r_{tr} = 500$, 400, 
300, 200, and 100 kpc.
As $r_{tr}$ decreases the current locus 
in the ($L_x/L_B,~r_{ex}/r_e$)-plane moves from the 
untruncated solution almost 
exactly through the observed correlation.
The filled square at the lower left is the locus of 
NGC 4472 if stellar mass loss creates all of the interstellar gas
(the $\Omega_b = 0$ solution).
\label{fig3}}

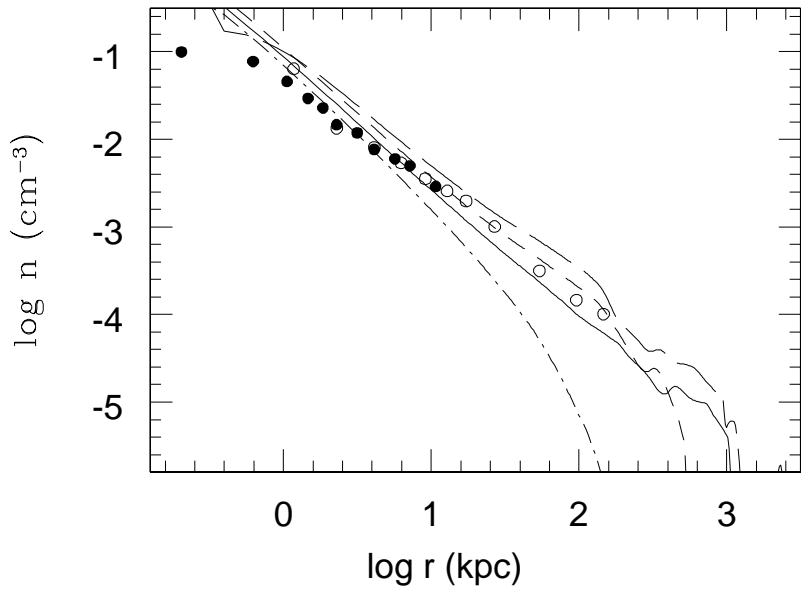
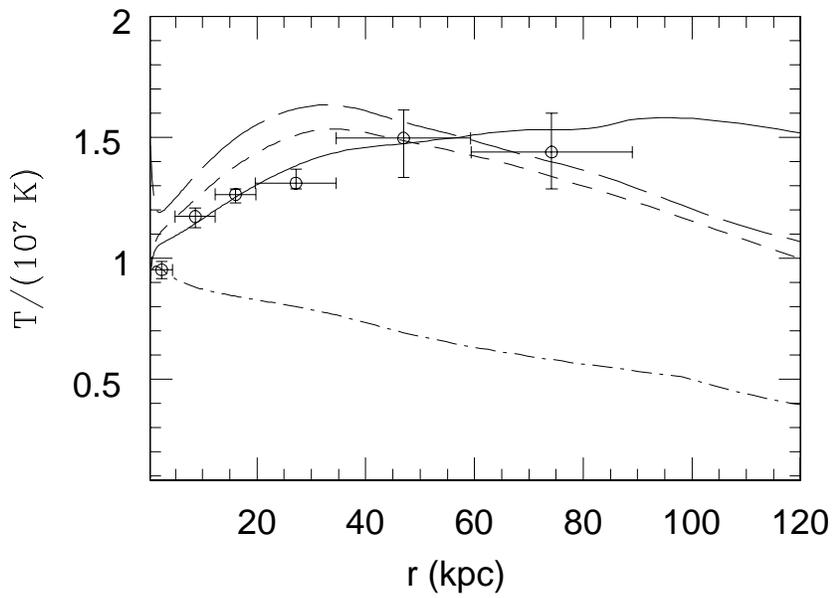
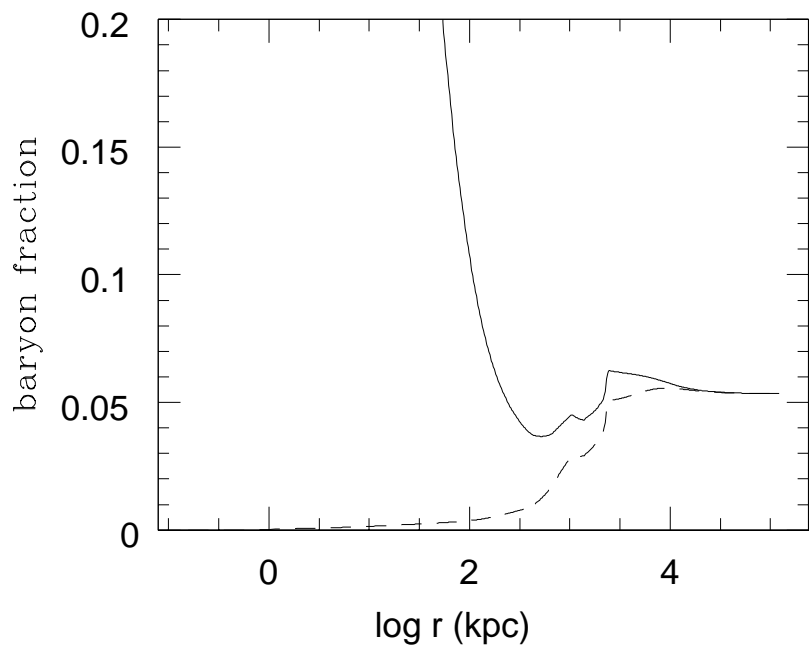
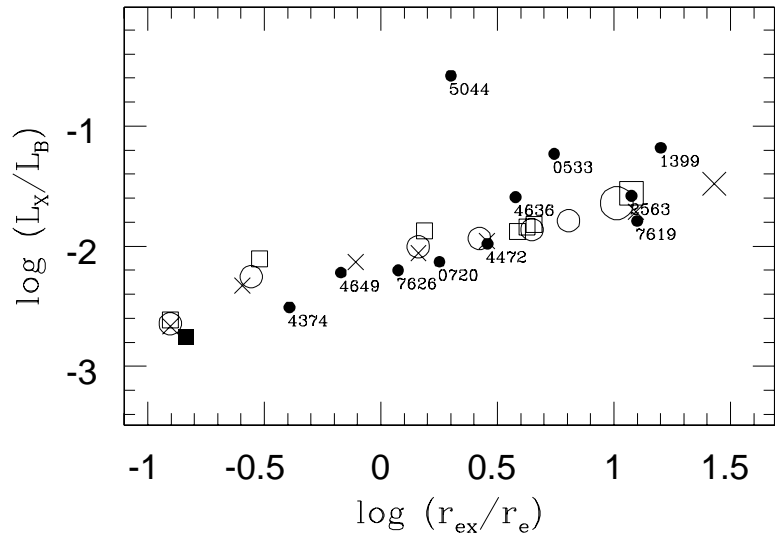
\end{document}